\documentclass{article}

\usepackage{arxiv}

\usepackage[utf8]{inputenc} 
\usepackage[T1]{fontenc}    
\usepackage{hyperref}       
\usepackage{url}            
\usepackage{booktabs}       
\usepackage{amsfonts}       
\usepackage{nicefrac}       
\usepackage{microtype}      
\usepackage{lipsum}
\usepackage{graphicx}
\graphicspath{ {./images/} }
\usepackage{subcaption}
\usepackage{multirow}
\usepackage{multicol}
\usepackage{bm}
\usepackage{amsmath}
\usepackage[font=small,labelfont=bf,figurename=Figure]{caption} 
\usepackage[capitalize,nameinlink]{cleveref}
\crefname{section}{Section}{Sections}
\crefname{subsection}{Subsection}{Subsections}

\title{Graph Neural Network-Based Predictive Modeling for Robotic Plaster Printing}

\author{
 Diego Machain Rivera \\
  Gramazio Kohler Research \\
  ETH Z{\"u}rich, Z{\"u}rich, Switzerland\\
  \texttt{machain@arch.ethz.ch} \\
   \And
 Selen Ercan Jenny \\
  Gramazio Kohler Research\\
  ETH Z{\"u}rich, Z{\"u}rich, Switzerland\\
  \texttt{ercan@arch.ethz.ch} \\
  \And
 Ping-Hsun Tsai \\
  Gramazio Kohler Research\\
  ETH Z{\"u}rich, Z{\"u}rich, Switzerland\\
  \texttt{tsai@arch.ethz.ch} \\
  \And
 Ena Lloret-Fritschi \\
  Academy of Architecture\\
  USI, Mendrisio, Switzerland\\
  \texttt{ena.lloret.fritschi@usi.ch} \\
  \And
 Luis Salamanca \\
  Swiss Data Science Center (SDSC)\\
  ETH Z{\"u}rich, Z{\"u}rich, Switzerland \\
  \texttt{luis.salamanca@sdsc.ethz.ch} \\
  \And
 Fernando Perez-Cruz \\
  Swiss Data Science Center (SDSC)\\
  ETH Z{\"u}rich, Z{\"u}rich, Switzerland \\
  \texttt{fernando.perezcruz@sdsc.ethz.ch} \\
  \And
 \hspace{-3mm}Konstantinos E. Tatsis \\
  Swiss Data Science Center (SDSC)\\
  ETH Z{\"u}rich, Z{\"u}rich, Switzerland \\
  \texttt{konstantinos.tatsis@sdsc.ethz.ch} \\
}

\begin{document}
\maketitle
\begin{abstract}
This work proposes a Graph Neural Network (GNN) modeling approach to predict the resulting surface from a particle based fabrication process. The latter consists of spray-based printing of cementitious plaster on a wall and is facilitated with the use of a robotic arm. The predictions are computed using the robotic arm trajectory features, such as position, velocity and direction, as well as the printing process parameters. The proposed approach, based on a particle reresentation of the wall domain and the end effector, allows for the adoption of a graph-based solution. The GNN model consists of an encoder-processor-decoder architecture and is trained using data from laboratory tests, while the hyperparameters are optimized by means of a Bayesian scheme. The aim of this model is to act as a simulator of the printing process, and ultimately used for the generation of the robotic arm trajectory and the optimization of the printing parameters, towards the materialization of an autonomous plastering process. The performance of the proposed model is assessed in terms of the prediction error against unseen ground truth data, which shows its generality in varied scenarios, as well as in comparison with the performance of an existing benchmark model. The results demonstrate a significant improvement over the benchmark model, with notably better performance and enhanced error scaling across prediction steps.
\end{abstract}


\section{Introduction}
\label{sec:Introduction}

Plastering, a craft that is as old as the history of building, is often considered one of the most important steps in building construction as it delivers the final finishing of the building structure (i.e. on brick or concrete walls). It requires years of training for craftspeople, and it is a time consuming and waste generating task, which is also extremely strenuous, affecting human safety and health. Throughout the history of architecture and construction, plasterwork used on interior walls and ceilings, as well as on facades, has played a functional role in providing durability, stability as well as visual, acoustic and light diffusing effects \cite{alma990069533620205503}. It is a multi-step process that combines spraying and troweling (smoothening) techniques, which involves removing a certain amount of the wet material, generating waste. 

\begin{figure}[h]
    \centering
    \includegraphics[width = 0.57\textwidth]{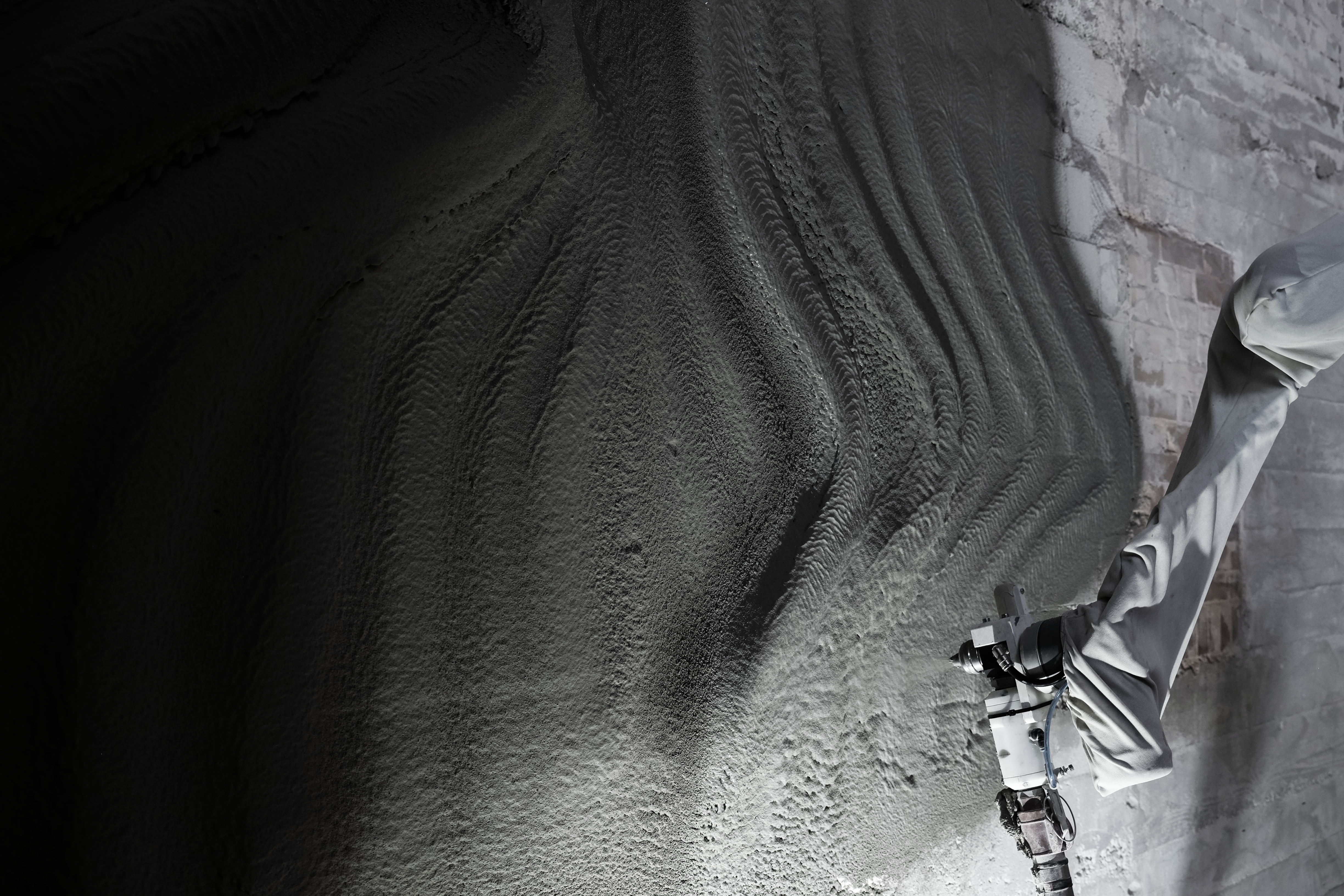}
    \caption{Robotic Plaster Spraying (RPS), introduces an additive-only, spray-based printing technique that can be applied directly onto a building structure.}
  \label{fig:RPS1}
\end{figure}

The technique presented in this paper, Robotic Plaster Spraying (RPS), introduces an additive-only, spray-based printing - resisting gravity - which enables iterative deposition of thin layers of plaster directly onto a building structure to create flat and bespoke surface finishing through digital control, without the need of any additional tools, support structures or molds, efficiently reducing the process into a single, spray-based step~\cite{ERCANJENNY2023104634}. This approach enables the application of the right amount of material “where necessary”, while eliminating the generation of unnecessary waste. 

\begin{figure}[h]
    \centering
    \includegraphics[width = 0.95\textwidth]{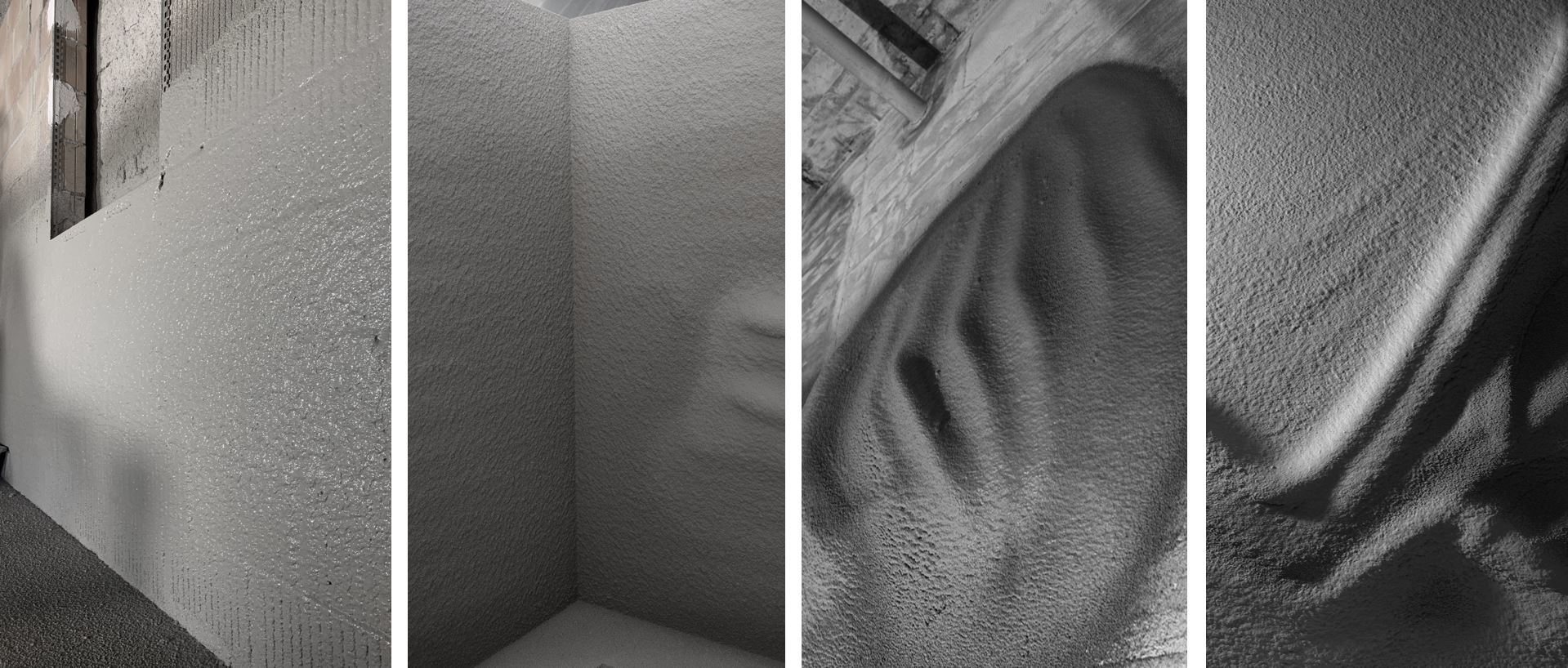}
    \caption{Examples of standardized and bespoke (custom) surfaces created with the proposed plaster printing process.}
  \label{fig:RPS2}
\end{figure}

RPS introduces an efficient mobile digital processing and fabrication system, whereby the material is robotically applied with the developed additive manufacturing technique, “adaptive thin-layer printing”, which is depicted in \cref{fig:RPS1}. The proposed solution is capable of achieving surface finishing at a speed of 2.4 min/m\textsuperscript{2}, for a total output of 200 m\textsuperscript{2}/day , and saving up to 20\% on material usage (i.e. for applied plaster), enabling time, and cost-efficient treatment of standard and bespoke walls and the creation of custom surface qualities, as shown in \cref{fig:RPS2}.

\begin{figure}
    \centering
    \includegraphics[width = 0.95\textwidth]{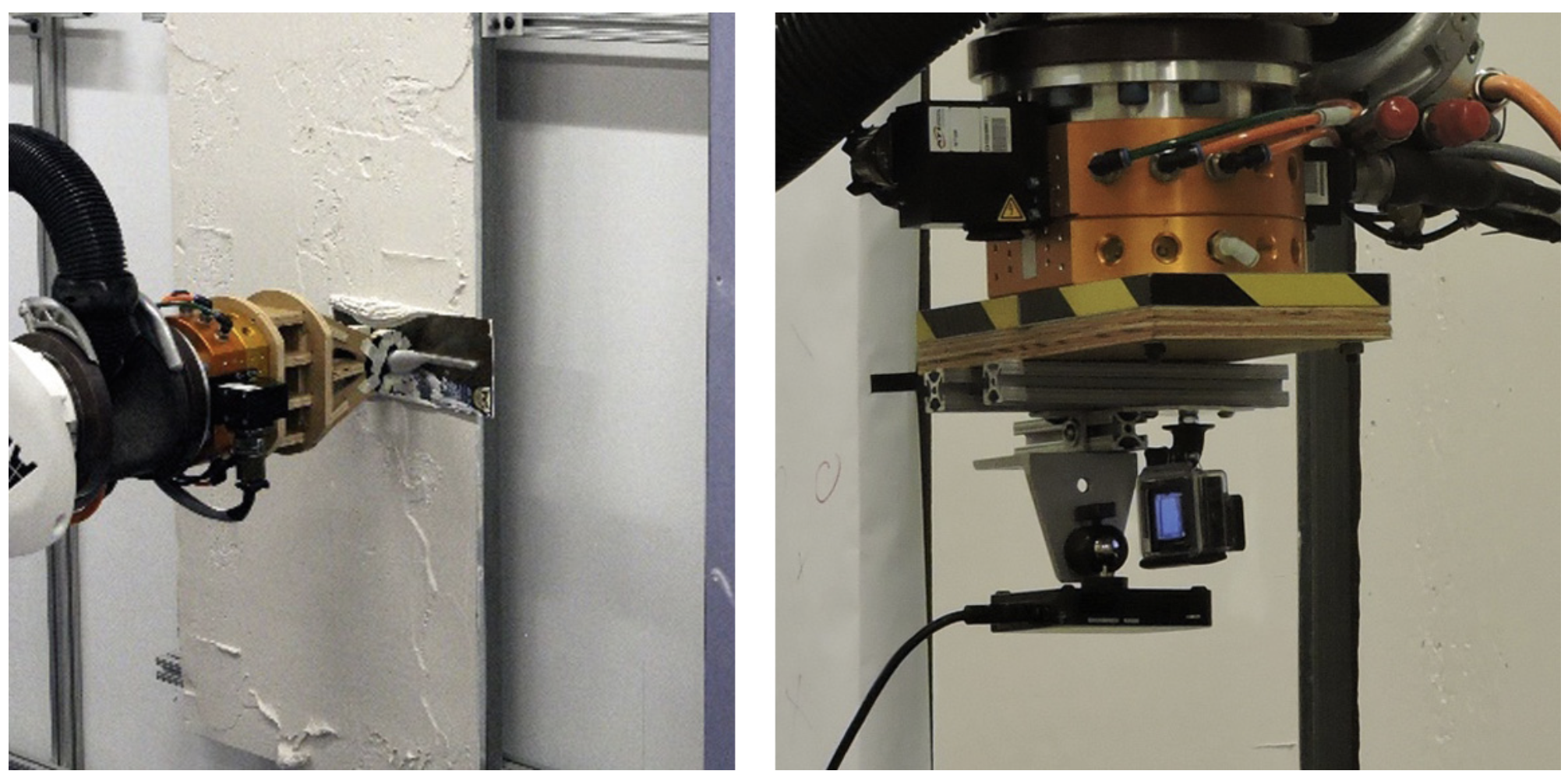}
    \caption{End-effector tools used by \cite{robotic_plastering}; left: troweling tool used in conventional plastering processes to smoothen the surfaces; right: camera-based inspection of the quality.}
    \label{fig:robotic_plaster}
\end{figure}

However, to be able to explore the constructive, aesthetic and performative potentials of combining sprayable materials, such as cementitious plaster, with a robotic arm, we need to anticipate, predict and visualize the spray outcome. In this regard, Machine Learning (ML)- and Deep Learning (DL)-based approaches could reveal untapped potentials by predicting possible failures and opening up the design space of similar sprayable materials that go beyond plaster and save resources, making building elements more efficient to produce. A relevant example is the image classification for robotic plastering with Convolutional Neural Networks (CNNs) \cite{robotic_plastering}. This approach detects and classifies discrepancies between the plaster design and material deposition on the wall surface in order to improve accuracy. The method relies on two end-effector tools, one to apply plaster and the other one to check the results with a camera, as illustrated in \cref{fig:robotic_plaster}. However, it does not allow the physical outcome of the plastering process to be predicted in order to inform the design process.

Some state of the art approaches that relate to the presented research include those applied in processes such as cutting foam: "Spatial Wire Cutting" research at ETH Zurich~\cite{20.500.11850/117537} investigated material- and fabrication process-related constraints, thereby making correlations between the controllable physical factors and responses of the process such as heat input, cutting speeds, resulting cutting forces, and wire shape. After the creation of a substantial database through iterative experiments, the method enabled the development of a data-driven design, simulation, and fabrication tool.

A similar approach was used for the modeling of a welding process with robots using Bayesian Networks and fusing data from different sources \cite{Kristiansen2007ModellingOT}. This project produced a model for process planning and control of an industrialized welding process, which required several parameters to be integrated into the model, ranging from material behaviour to the speed and the force of welding. The project proposed a simulation approach that encapsulates data from empirical, analytical, and operators’ knowledge from welding processes, all encoded together and fed into a machine learning algorithm, which was iteratively tested and updated. The final result was a data-driven simulation tool for process-planning in welding operations. Such projects suggest that it may not be necessary to fully understand material behaviour at the granular or micro level. Instead, they demonstrate the possibility of managing material behaviour in response to the needs of the operation system in use, without needing a complete, detailed and precise description of the underlying physics. 

One of the main challenges in the domain of manufacturing and construction is related to the demand for increasingly complex and high-quality products, in terms of design principles, standardization and quality control. Within this context, (ML) models rise to play a critical role as they are able to provide effective digital means of quality control, process optimization, modeling of complex systems, and energy management. This digitalization layer has been investigated for the planning and simulation of 3D printing applications with the use of Building Information Modeling (BIM) \cite{Paolini2019}. Furthermore, the use of ML and DL techniquesin the additive manufacturing domain has been explored in order to optimize design and production processes. An overview of the state-of-the-art ML methods used in this context is presented in \cite{Wang2020}, while the work of \cite{Mu2024} is focused on the development of generative models for digital twin applications in the manufacturing of metallic parts. The use of machine learning models has been also used for the detection of defects and the evolution of materials during the manufacturing processes, with a review presented in \cite{Parsazadeh2023}. Lastly, a recent review of ML methods applied for the shift towards more efficient, sustainable and automated construction and additive manufacturing processes is presented in \cite{Qin2022}.

Despite the recent developments in digital manufacturing, architecture and construction, the use of ML and DL solutions for design and fabrication processes still remains relatively little explored. A pioneering example for facilitating a data-driven approach to quantifying the quality of the surfaces of building elements, i.e. roughness, smoothness, etc., with expert-informed feature generation can be seen in~\cite{Frangez2022, Frangez_PhD_2022}. ML-based approaches have the potential to predict and visualize what we can actually build, and existing literature points towards the potentials and constraints~\cite{10.1145/3485114.3485122, AIR, 10.1007/978-3-319-92294-2_26}. However, they mainly target the usage of modular systems such as wood assembly~\cite{APOLINARSKA2021103569} or measurement on a series of 3D printed panels to define the optimal surfaces for acoustics~\cite{doi:10.1177/1351010X20986901}. Based on the available literature, we can conclude that ML-based approaches have not been yet explored to predict and visualize complex-to-simulate material behaviour, such as concrete or cementitious plaster.

In this study, we propose a new method for predicting the thickness of printed plaster from the trajectory and operational parameters of the robot. The method involves obtaining thousands of individual predictions from some trajectory steps, with multiple trajectory steps forming a layer. The goal is to get predictions of the thickness for future layers without the need to implement them in real life to inform the design process. Our approach builds on previous work conducted within the Robotic Plaster Spraying (RPS) project, which focused on planning the design and fabrication process before building up the layers with cementitious plaster and predicting the thickness of the material printed on the wall based on the fabrication parameters. These parameters included the vertical distance of points on the wall surface to the spraying (printing) path projected to the surface, the end-effector distance to the transformed mesh, the velocity of the trajectory, and the layer number. The prediction was based on a nonlinear regression model \cite{plastering}, delivering layer by layer thickness predictions.


\section{Problem Description}
\label{sec:problem-description}

The recent advancements in construction robotics have introduced autonomous plastering processes facilitated with additive manufacturing techniques based on robotic arms. As shown with Robotic Plaster Spraying (RPS), it is possible to expand the design space of building surfaces by repeatedly printing thin layers of plaster to build up complex volumetric formations or textural patterns. Although such a plastering process relies on sensing and control units to maintain a high degree of control, there are still features that can offer an additional level of control and versatility to the outcome.

\begin{figure}[h]
    \centering
    \includegraphics[width = 0.65\textwidth, page = 16]{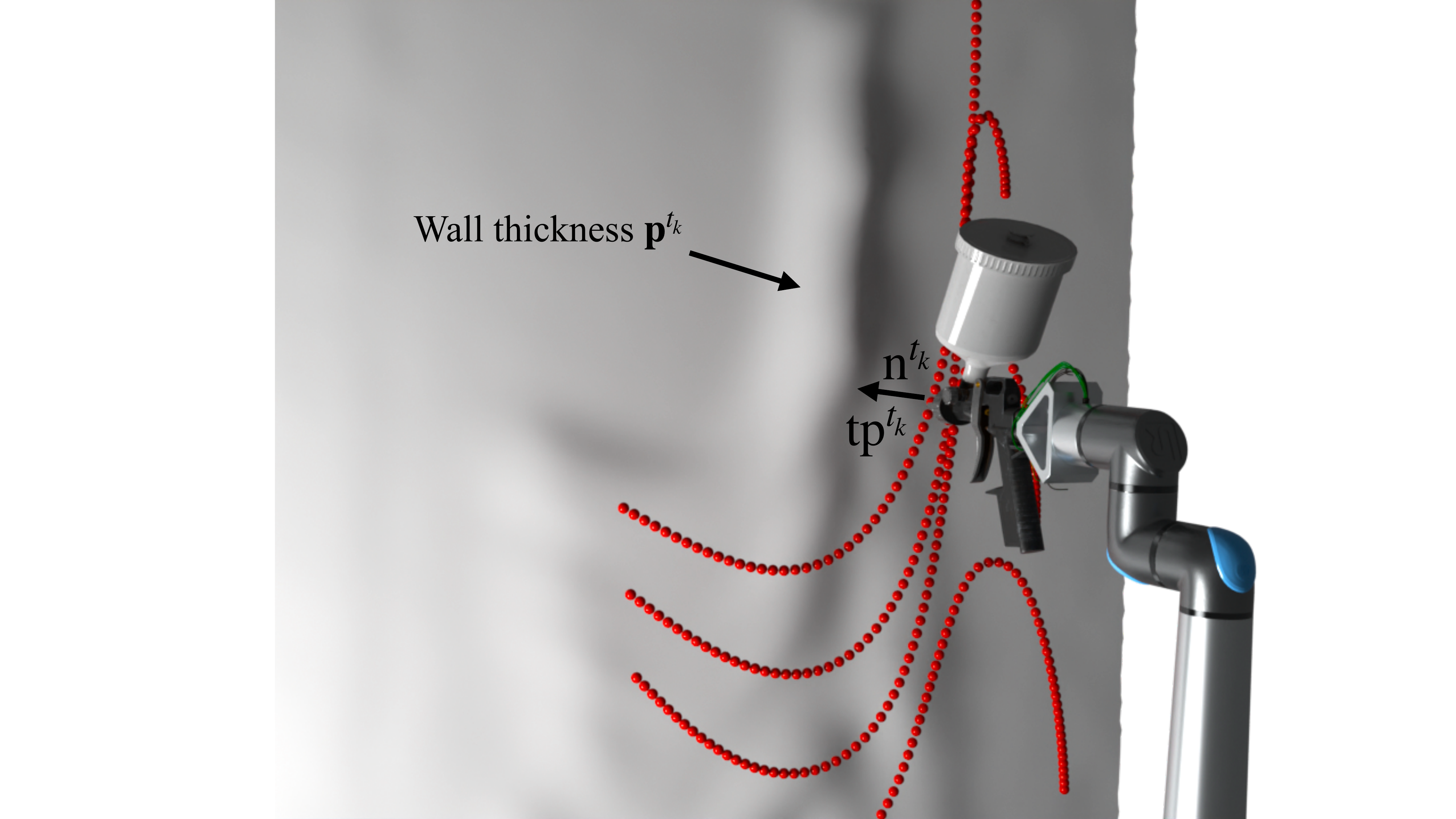}
    \caption{Graphical representation of the plastering process; the wall thickness at time step $t_k$ is represented by the point cloud $\mathbf{p}^{t_k}$; the operational point of the spraying gun is described by the working pressure $P^{t_k}$, the position $\mathbf{tp}^{t_k}$, the direction of spray-based printing $\mathbf{n}^{t_k}$ and the velocity $\mathbf{u}^{t_k}$.}
    \label{img:spraying_gun}
\end{figure}

Within this context, the robotic plaster printing process is described in this section from a
predictive modeling point of view, where the task involves the prediction of the wall thickness using the data acquired by a robotic arm with a spray gun at the end effector. In order to generate such predictions, the underlying physics between the working parameters of the robotic arm and the end-effector should be learned. The operation of the spray gun is characterized by a number of parameters that include the working pressure $P^{\,t_k}\in\mathbb{R}$, the trajectory position $\mathbf{tp}^{t_k}\in\mathbb{R}^3$ of the end effector in each time step $t_k$, for $k=0,1,\ldots,T$, along with the associated velocity $\mathbf{u}^{t_k}\in\mathbb{R}^3$ and the spraying (spray-based printing) direction $\mathbf{n}^{t_k}\in\mathbb{R}^3$. The plastering process is based on the printing of multiple thin layers, each of which is produced once the spraying gun has traveled over a trajectory $\mathbf{T} = \left[\mathbf{tp}^{t_{0}}, \mathbf{tp}^{t_{1}},..., \mathbf{tp}^{t_{T}}\right]$ that consists of $T$ time steps. The thickness of the wall is represented by a point cloud $\bold{p}^{t_k}\in\mathbb{R}^{N\times 3}$ that describes the position of the $N$ particles at each time step $t_k$. These particles are structured in an orthogonal grid that covers the entire domain to be plastered, while the number of these particles $N$ is kept constant during the printing and modeling phase. Moreover, their in-plane position is always fixed and only the out-of-plane coordinate is allowed to change, which essentially represents the wall thickness. All variables of the problem are graphically depicted in \cref{img:spraying_gun}, in which the red points represent the trajectory positions of the end-effector.

Based on the previous domain representation, the aim of the predictive model can be described as follows. Given some input trajectory position $\mathbf{tp}^{t_k}$ at each time instant $t_k$ and the operational parameters of the spray gun and the point cloud $\mathbf{p}^{t_{k}}$, the model operates recursively to output the point cloud $\mathbf{p}^{t_{k+1}}\in \mathbb{R}^{N\times 3}$, that describes the position of the particles at the next step, namely $t_{k+1}$. In a nutshell, the model is just calculating the change of the wall thickness from $t_k$ to $t_{k+1}$, by tracking the evolution of the particles in the out-of-plane direction.

\subsection{Data}
\label{subsec:data}

The data for the plaster printing process was collected under laboratory conditions by the project team at the Gramazio Kohler Research group at ETH Z{\"u}rich. The volumetric formations produced during the experiments consist of several layers, each printed by a 6-degree-of-freedom (6-DOF) robotic arm, which follows multiple trajectories while depositing layers of cementitious plaster on a wall. The collected dataset contains discretized and sparse in time values of the following problem parameters for a number of printed layers:

\begin{itemize}
  \item Trajectory positions of the robotic arm $\mathbf{tp}^{t_k}$
  \item Positions of the wall particles $\mathbf{p}^{t_k}$
  \item Velocity of the robotic arm $\mathbf{u}^{t_k}$
  \item Printing directions of the end effector $\mathbf{n}^{t_k}$
  \item Air pressure of the plaster printing (spray) gun $P^{\,t_k}$
\end{itemize}

The data is acquired from five different experiments, which are carried out under different conditions. The domain of these experiments covers an area of around 6.40 m\textsuperscript{2} and is represented by 11.000 particles. The main operational parameters, namely the velocity of the robotic arm and the distance of the spray gun from the wall, are not kept fixed in time and across the different tests. Their statistics are summarized in \cref{tab:data-statistics}.

\begin{table}[h]
    \centering
    \small{
    \begin{tabular}{@{}cccccccccccc@{}}
        \toprule
        & \multirow{2}{*}{Experiment} & \multicolumn{4}{c}{Velocity [m/s]} & & \multicolumn{4}{c}{Distance to wall [mm]} &\\
        \cmidrule(l){3-6}\cmidrule(l){8-11}
        & & mean & std & max & min & & mean & std & max & min & \\ 
        \midrule
        & 1 & 0.81 & 0.26 & 1.00 & 0.10 & 
          & 445.1 & 20.5 & 506.4 & 405.6 & \\
        & 2 & 0.72 & 0.25 & 1.00 & 0.10 & 
          & 336.3 & 35.9 & 426.5 & 259.2 & \\
        & 3 & 0.55 & 0.28 & 1.00 & 0.10 & 
          & 228.6 & 33.1 & 308.0 & 170.6 & \\
        & 4 & 0.76 & 0.16 & 1.00 & 0.15 & 
          & 388.0 & 15.0 & 417.9 & 352.9 & \\
        & 5 & 0.65 & 0.20 & 1.00 & 0.10 & 
          & 333.4 & 21.2 & 383.5 & 296.7 & \\
        \bottomrule
    \end{tabular}
    }
    \caption{Statistics of the spray gun velocity and distance of the robotic arm from the wall during the five experiments contained in the collected dataset}
    \label{tab:data-statistics}
\end{table}

\begin{figure}[h]
    \centering
    \begin{subfigure}{0.35\textwidth}
        \includegraphics[width=\linewidth,trim={5cm 5.8cm 0 0},clip]{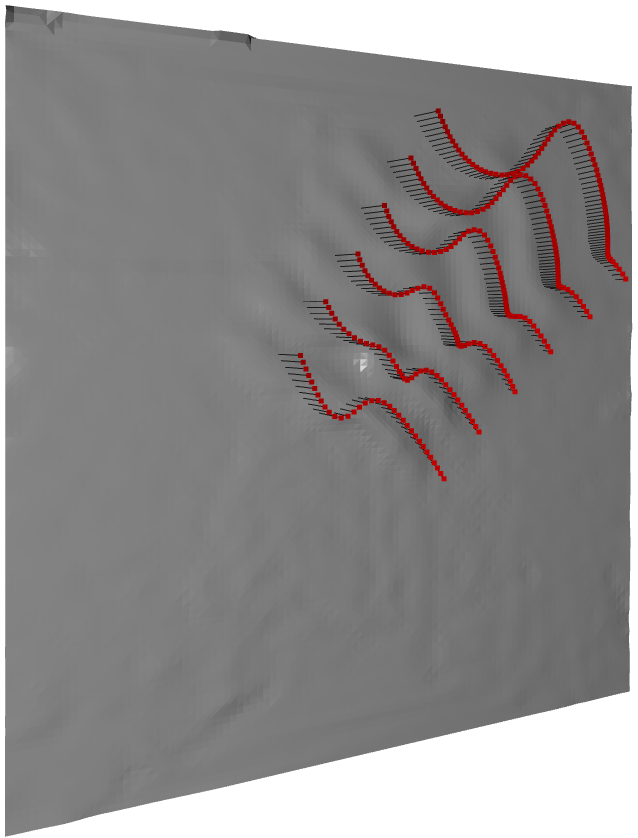}
        \label{fig:sub2}
    \end{subfigure}
    \hspace{10mm}
    \begin{subfigure}{0.35\textwidth}
        \includegraphics[width=\linewidth,trim={5cm 5.8cm 0 0},clip]{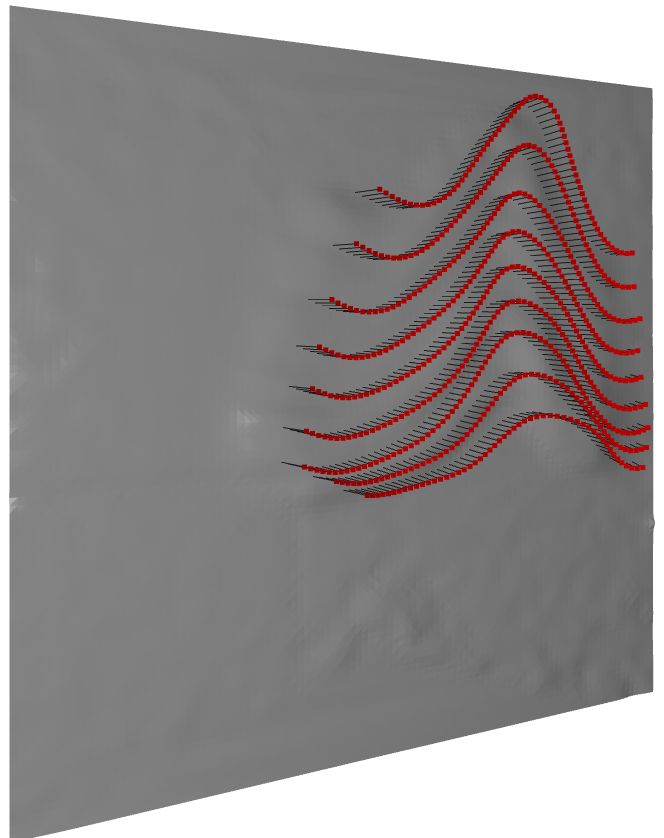}
        \label{fig:sub3}
    \end{subfigure}
    \caption{Illustration of a plaster pattern along with the spray-based printing trajectories, whose positions $\mathbf{tp}^{t_k}$  at each step are indicated by red points; the small vectors on each trajectory point represent the printing direction.}
    \label{fig:data_images}
\end{figure}

The point clouds $\mathbf{p}^{t_k}$ that represent the state of the wall in terms of the thickness pattern are captured with the aid of an RGB-D sensor. The scans were generated layer by layer, meaning that the end effector had passed through the entire trajectory at least once, before the wall thickness was recorded. Some representative examples of the trajectories and the wall pattern are shown in \cref{fig:data_images}. The small vectors on each trajectory point represent the printing direction, where the variability can be observed at the bottom left corner of the leftmost figure. Lastly, due to the sparsity of measurements in time, an additional step of data augmentation was implemented to improve the training of the predictive model. 

\subsection{Data augmentation}
\label{subsec:data-augmentation}

The collection of data during the plaster printing process is limited by a number of practical constraints that cannot be easily overcome. This implies that although the trajectories of the spray gun are fully recorded over time, the point cloud that represents the wall thickness can be only sparsely measured over time, thus resulting to a dataset that cannot be used directly for the training of an one-step ahead predictor. It should be clarified here that a single time step of the plaster printing process corresponds to the transition of the spray gun from a trajectory position $\mathbf{tp}^{t_k}$ to the next one $\mathbf{tp}^{t_{k+1}}$. As such, the original dataset is herein augmented, as described in the remaining of this section, in order to deliver the so-called augmented dataset, as described in \cref{tab:data-augmentation}. It is observed that despite the fact that all experiments consist of 16 layers, the original dataset contains only a few point clouds, ranging from 2 to 4. Upon augmentation, the total number of point clouds is increased to a few thousands, a number that corresponds to a point cloud per trajectory step for each layer.

\begin{table}[h]
    \centering
    \small{
    \begin{tabular}{@{}cccccccccc@{}}
        \toprule
        & \multirow{3}{*}{Experiment} & \multirow{3}{*}{Layers} & & \multicolumn{2}{c}{Original dataset} & & \multicolumn{2}{c}{Augmented dataset} & \\\cmidrule(lr){5-6}\cmidrule(lr){8-9}
        & & & & \multicolumn{1}{c}{Trajectory} & \multicolumn{1}{c}{Point} & & \multicolumn{1}{c}{Trajectory} & \multicolumn{1}{c}{Point} & \\
        & & & & \multicolumn{1}{c}{Steps} & \multicolumn{1}{c}{Clouds} & & \multicolumn{1}{c}{Steps} & \multicolumn{1}{c}{Clouds} & \\ \midrule
        & 1 & 16 & & 415 & 3 & & 415 & 6640 & \\
        & 2 & 16 & & 391 & 4 & & 391 & 6256 & \\
        & 3 & 16 & & 314 & 3 & & 314 & 5024 & \\
        & 4 & 16 & & 472 & 3 & & 472 & 7552 & \\
        & 5 & 16 & & 417 & 2 & & 417 & 6672 & \\
        \bottomrule
    \end{tabular}}
    \caption{Comparison of the dataset before and after the augmentation}
    \label{tab:data-augmentation}
\end{table}

\subsubsection{Inter-layer augmentation}
\label{subsubsec:inter-layer-augmentation}

One of the main limitations during the collection of data is related to the fact that the measurement of the wall thickness requires the interruption of the printing process. This results in a process that is practically difficult to implement in parallel with the printing, which calls for the collection of data points that are sparsely distributed in time. As such, the point clouds that represent the wall thickness were sparsely recorded between the different spraying layers, rather than consecutively. For instance, in one experiment the data was collected before the spray-based printing of layers 0, 1, 10 and 20, while in another experiment the wall thickness was measured before the spraying of layers 0, 1, 2, 4, 6 and 8.

To circumvent this limitation of the data acquisition process, a linear interpolation was used to calculate the thickness before the printing of each layer. This was validated by investigating the effect of spraying between consecutive layers, which was well approximated by such an interpolation scheme. Therefore, the original dataset was augmented by calculating the position difference for each particle between the recorded layers and thereafter generating the interpolated positions to fill in the missing layers information. It should be noted that the remaining printing parameters, such as velocity, pressure and trajectory positions, are kept fixed in between the layers.

\subsubsection{Inter-step augmentation}
\label{subsubsec:inter-step-augmentation}

\begin{figure}
    \centering
    \includegraphics[width=0.45\textwidth]{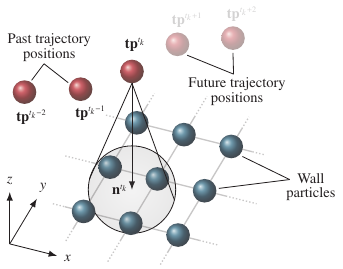}
    \caption{Illustration of the cone created from each trajectory point in alignment with the spray-based printing direction $\mathbf{n}^{t_k}$ for the calculation of the influence area in the inter-step augmentation process}
    \label{fig:inter-step-augmentation}
\end{figure}

The second augmentation step aims at delivering the wall thickness information at the intermediate time steps of a printing trajectory. To do so, a cone $\mathcal{C}^{t_k}$ is used at each time instant to quantify the effect of the printing process on the wall, as shown in \cref{fig:inter-step-augmentation}. The vertex of the cone denotes the spraying position $\bold{tp}^{t_{k}}$ and the direction of the vertex indicates the printing direction $\mathbf{n}^{t_k}$. Therefore, the augmentation process captures the effect that only the particles within the cone $\mathcal{C}^{t_k}$ are the only ones whose position is affected by the spraying gun. The value for the base radius of the cone, which is half the distance from the printing point to the wall, is based on the values reported in existing literature \cite{plastering}.

Following the illustration in Fig. \cref{fig:inter-step-augmentation}, the steps of this augmentation process work as follows: for each layer $l$, the position $\mathbf{p}^{l,t_0}$ of the wall particles at step $t_0$ is set equal to the particles position at the end of the last step $t_T$ of the previous layer, namely $\mathbf{p}^{l-1,t_T}$. For the calculation of the particles position $\mathbf{p}^{l,t_1}$ at step $t_1$ for layer $l$, the area of influence of the spray gun is calculated according to the cone $\mathcal{C}^{t_1}$, as described in the previous paragraph, and the position of all the particles falling within that area, which is denoted by $\bar{\mathbf{p}}_i^{l,t_1}\in\mathcal{C}^{t_1}$, is updated and set equal to $\mathbf{p}_i^{l+1,t_1}\in\mathcal{C}^{t_1}$. This process is iteratively repeated for all trajectory positions $\mathbf{tp}^{t_k}$ of the spray gun during the printing of the $l$-th layer, thus delivering the sequence of point clouds $\mathbf{M}^l = \left[\bar{\mathbf{p}}^{l,t_0}, \ \bar{\mathbf{p}}^{l,t_1},\ \ldots ,\  \bar{\mathbf{p}}^{l,t_T} \right]$, where $T$ denotes the total number of steps. This augmentation step results in a total number of point clouds equal to $\sum_{l=1}^L T_{l} \times l$, each of which is of size $N \times 3$. The total number of point clouds for each experiment before and after the augmentation is shown in \cref{tab:data-augmentation}.

\section{Methodology}
\label{sec:methodology}

The goal of this section is to formulate a Graph Neural Network (GNN) model for the prediction of the wall thickness change during the plaster printing process, on the basis of the problem description presented in \cref{sec:problem-description}. To do so, the particles of the wall are represented by graph nodes, which interact with the spray gun through the end effector. The latter is also represented by a single particle, whose position is determined by the robotic arm and as such, its motion in space is dictated by the velocity $\mathbf{u}^{t_k}$ of the spray gun. The state of the volumetric formation at each time step can be completely described by the position $\mathbf{p}^{t_k}$ of the wall particles with respect to the initial reference wall, the velocity $\mathbf{u}^{t_k}$ and direction $\mathbf{n}^{t_k}$ of the spray gun, as well as the trajectory position $\mathbf{tp}^{t_k}$ of the end effector. Within this context, the aim of the GNN model is to learn the underlying mechanics of the printing process as a map between the printing parameters and the wall thickness. Moreover, such a mapping will be established in a recursive form, which predicts the change of position $\Delta\mathbf{p}^{t_{k+1}}$ of the wall particles from the initial position $\mathbf{p}^{t_k}$ at step $t_k$ to the one at step $t_{k+1}$, denoted by $\mathbf{p}^{t_{k+1}}$, given the values of the end effector parameters at step $t_k$.

\begin{figure}[h]
    \centering
    \begin{subfigure}{0.32\textwidth}
        \includegraphics[width=\textwidth]{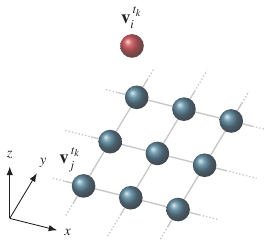}
        \caption{}
    \end{subfigure}
    \begin{subfigure}{0.32\textwidth}
        \includegraphics[width=\textwidth]{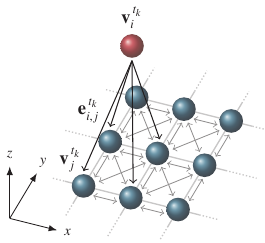}
        \caption{}
    \end{subfigure}
    \begin{subfigure}{0.32\textwidth}
        \includegraphics[width=\textwidth]{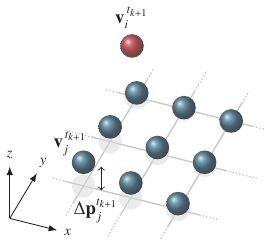}
        \caption{}
    \end{subfigure}
    \caption{Schematic representation of the graph-based modeling approach; (a) at each time step $t_k$ the input of the model consists of the node features $\mathbf{v}^{t_k}$ that describe the state of the domain; (b) the graph is created through the edges $\mathbf{e}_{i,j}$ that describe the particle connectivities; (c) the wall particle position changes $\Delta \mathbf{p}^{t_{k+1}}$ are predicted from the graph model.}
    \label{fig:general_methodology}
\end{figure}

The model divides the type of particles into wall particles and one end effector particle. Both of them are represented as nodes in the graph model however, the former set of particles is described by the position $\mathbf{p}^{t_k}$ as nodal feature, while the latter is further characterized by the velocity $\mathbf{v}^{t_{k}}$ and the printing direction $\mathbf{n}^{t_{k}}$. It should be noted that the position of the end effector is denoted by $\mathbf{tp}^{t_{k}}$, which is determined by the trajectory of the robotic arm. A schematic representation of graph-based modeling approach is presented in \cref{fig:general_methodology}. The starting point at each time step is the node features of the wall and end effector particles, which enable the construction of the graph by establishing the connectivities among the particles. The set of node $\mathbf{v}^{t_k}$ and edge $\mathbf{e}^{t_k}$ features is subsequently given as input to the graph model for the prediction of the change of position $\Delta\mathbf{p}^{t_{k+1}}$ for each wall particle

\begin{equation}
    \Delta\mathbf{p}^{t_{k+1}} = \mathcal{M}_{\theta}\left(\mathbf{v}^{t_k}, \mathbf{e}^{t_k}\right)
    \label{eq:gnn-prediction}
\end{equation}

where $\bm{\theta}\in\mathbb{R}^{n_{\theta}}$ contains all the trainable model parameters. The model $\mathcal{M}_{\theta}$ is divided into three Graph Network blocks and follows an \textit{encode - process - decode} structure, which is more thoroughly described in \cref{subsec:model-structure}.

\subsection{Graphs}
\label{subsec:graphs}

A graph is defined as a 3-tuple $G=(V,E,\bold{q})$, where $V=\{\bold{v}_{i}\}_{i=1:N_{\mathrm{v}}}$ denotes the set of nodes, with $\bold{v}_{i}$ being a vector that contains the attributes of the $i-$th node among the $N_{\mathrm{v}}$ nodes and $E=\{(\bold{e}_{j}, r_{j}, s_{j})\}_{j=1:N_{\mathrm{e}}}$ is the set of all the $N_{\mathrm{e}}$ edges. The edge attributes are denoted by $\bold{e}_{j}$, while $r_j$ and $s_j$ designate the indices of the receiver and sender nodes, respectively. Lastly, $\mathbf{q}$ denotes a vector of global graph features. This same graph structure is used in Graph Neural Networks (GNN), which constitute a concatenation of Graph Network (GN) blocks \cite{battaglia2018relational}. In this work, GN blocks are used along with their update and aggregation functions, which define the main operations being applied between the nodes and the edges of the graph. These operations are compactly described as follows

\begin{subequations}
    \begin{align}
        \text{Edge update:} & & \bold{e}'_{j}&=\phi ^{\mathrm{e}}(\bold{e}_{j}, \bold{v}_{r_{j}}, \bold{v}_{s_{j}})
        \label{eq:edge-update}\\[1.5mm]
        \text{Node update:} & & \bold{v}'_{i}&=\phi ^{\mathrm{v}}(\bold{\bar{e}}'_{i}, \bold{v}_{i})
        \label{eq:node-update}\\[0.75mm]
        \text{Aggregation:} & & \bold{\bar{e}}'_{i}&=\rho^{\mathrm{e \rightarrow v}}(E'_{i}) = \sum_{j:r_{j}=i}\bold{e}'_{j}.
        \label{eq:edge-aggregation}
    \end{align}
    \label{eq:graph-operations}
\end{subequations}
\noindent
where $\phi^{\mathrm{e}}$ and $\phi^{\mathrm{v}}$ denote the edge and node update functions respectively, while $\rho^{\mathrm{e}\rightarrow \mathrm{v}}$ is the edge aggregation function. All three functions are unknown and learned, through the training phase of the model, with the aim of reflecting the interactions among the features of the GN.

\subsection{Model structure}
\label{subsec:model-structure}

\begin{figure}[h]
    \centering
        \includegraphics[width = 0.67\textwidth]{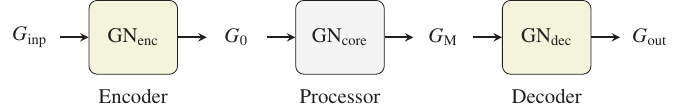}
        \caption{Schematic representation of the graph model consisting of multiple graph blocks; the input graph $G_{\text{inp}}$ is passed through an encoder to create the graph $G_{0}$; the latter is fed into the processor, which consists of $M$ graph blocks and generates $G_{M}$; the last step is a decoder that delivers the final graph $G_{\text{out}}$, which contains the change of position $\Delta \mathbf{p}_i$ for each one of the wall particles.}
        \label{fig:general_structure}
\end{figure}

The graph model consists of an \textit{encode - process - decode} layout, the three main GN blocks as described in the architecture presented in \cite{hamrick2018relational}. A schematic representation of the model layout is shown in \cref{fig:general_structure}. At each time step, the graph representation of the problem is constructed by populating the node attributes and building the connectivity between nodes. The latter is thoroughly described in the next subsection. The encoder block, which is denoted by GN\textsubscript{enc}, receives as input the graph representation of the problem, denoted by $G_{\text{inp}}$, and transforms it into a latent representation $G_{0}$. The processor GN\textsubscript{core} = $\text{GN}_{1} \circ  \text{GN}_{2} \circ \ldots \circ \text{GN}_{M}$ is a concatenation of $M$ blocks, which transfer information among the nodes using $M$ message-passing steps, and deliver the updated latent representation $G_M$. Lastly, the decoder block GN\textsubscript{dec} transforms the latent graph into $G_{\textrm{out}}$, which delivers as output, in the form of node attributes, the change of position $\Delta \mathbf{p}$ of the wall particles. The detailed implementation steps are documented in \cref{sub:implementation}.

\subsection{Connectivity}

The physics of the plaster printing process are modeled through the interaction of the end effector with the particles of the wall and the interaction among the wall particles themselves. These interactions are taken into account through the connectivity of the nodes, which is encoded in the edge attributes $\mathbf{e}_k$. As such, two different types of connectivities are considered, with the first one encoding the interaction of the end effector with the wall particles and the second one being responsible for the interaction among the wall particles.

\subsubsection{Connectivity between end effector and wall particles}

\begin{figure}[h]
    \centering
    \begin{subfigure}{0.32\textwidth}
        \includegraphics[width=\textwidth]{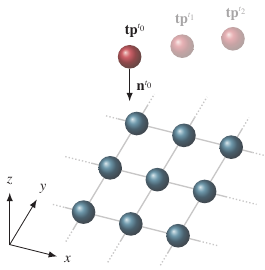}
        \caption{}
    \end{subfigure}
    \begin{subfigure}{0.32\textwidth}
        \includegraphics[width=\textwidth]{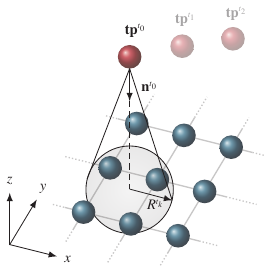}
        \caption{}
    \end{subfigure}
    \begin{subfigure}{0.32\textwidth}
        \includegraphics[width=\textwidth]{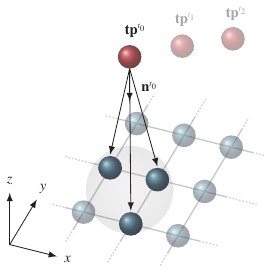}
        \caption{}
    \end{subfigure}
    \caption{Process of connecting the end effector particle with the wall particles; (a) projection of the printing direction into the wall plane; (b) selection of the circle radius $R^{t_k}$ based on the distance of the end effector particle to the wall; (c) connection of the particles that lie inside the cone to the end effector particle.}
    \label{fig:trajectory-wall-con}
\end{figure}

The connectivity of the wall particles with the end effector particle is dependent on the position of the latter along the spraying trajectory. As such, this connectivity is updated at each time step $t_k$ according to the following steps: the influence area of the spray gun is calculated at the position $\mathbf{tp}^{t_k}$ of the end effector at time step $t_k$. This is represented by a cone whose vertex is located at the trajectory point $\mathbf{tp}^{t_k}$ while its axis is aligned with the printing direction $\mathbf{n}^{t_k}$. The radius $R^{\,t_k}$ of the basis, which is located on the wall and defines the influence area of the spray gun, is proportional to the distance of the spray gun from the wall, which is essentially the distance of the trajectory point $\mathbf{tp}^{t_k}$ from the wall. This effect has been studied and quantified in \cite{plastering}. 
An one-way connection is created between the end effector particle and each particle within its area of influence as presented in \cref{fig:trajectory-wall-con}.

\subsubsection{Connectivity among wall particles}

In contrast to the connectivity between the end effector particle and the wall particles, which is used to model the direct effect of the spray gun, the connectivity among the wall particles is used in order to control the smoothness of the plaster printing process. This is achieved by adopting a nearest neighbour \cite{k-nearest} connectivity among all wall particles, which implies a two-ways connection for nodes whose in-plane distance is less than a threshold $R$. The degree of connectivity is inversely proportional to the nodes distance and the threshold value is treated as a hyperparameter.

\subsection{Implementation}
\label{sub:implementation}

This subsection offers a detailed description of the input-output representation and the building blocks of the model, as schematically presented in \cref{fig:general_structure}. The input of the model is constructed at each time step of the printing process and consists of the specification of the node and edge attributes. These are subsequently encoded into a latent space, using the node and edge encoders respectively. The processor applies message passing through the detailed \cref{eq:graph-operations} on the latent representation, and lastly, the decoder is applied to the node attributes, thus delivering the target output.

\subsubsection{Input-output representation}
\label{subsub:input-output-representation}

For the instantiation of the input graph $G_{\text{inp}}$, the particle positions $\mathbf{p}_i^{t_k}$ are used as node attributes $\bold{v}_i$. In order to distinguish the end effector from the wall thickness representation, a particle type is also included in the node attributes, which is 0 for the wall particles and 1 for the end effector particle. The latter comprises additional attributes related to the trajectory, such as the velocity $\mathbf{v}^{t_k}$ and the printing direction $\mathbf{n}^{t_k}$. The edge attributes $\bold{e}_{j}$ consist of the relative position $\mathbf{r}_j^{t_k} = \mathbf{p}_{r_j}^{t_k} - \mathbf{p}_{s_j}^{t_k}$ between connected particles, where subscripts $r_j$ and $s_j$ denote the receiver and sender nodes of the $j$-th edge, as well as the Euclidean distance $d_{\,\mathrm{j}}^{\,t_k} = ||\mathbf{r}_{j}^{t_k}||_{2}$ of the relative position, resulting in a total of 4 attributes. Lastly, the output node attributes, delivered by $G_\mathrm{out}$, contain the change of position $\Delta\mathbf{p}_{i}^{t_{k+1}} = \mathbf{p}_i^{t_{k+1}} - \mathbf{p}_i^{t_k}$ of the wall particles.

\subsubsection{Encoder}
\label{subsub:encoder}

The \textit{encoder} is responsible for the transformation of the particle-based representation of the wall-printer system  $G_{\mathrm{inp}}$ into the first latent graph $G_{0}$. In this latent space, the node attributes are encoded into a latent space through the encoder $\epsilon^{\mathrm{v}}: \mathcal{V}\to\mathcal{V}_0$, so that
$\bold{v}_{i,0} = \epsilon^{\mathrm{v}}(\bold{v}_{i})$. 
Similarly, the edge attributes are mapped to a latent space according to 
$\bold{e}_{j,0}=\epsilon^{\mathrm{e}}(\bold{e}_{j})$, where $\epsilon^{\mathrm{e}}:\mathcal{E}\to\mathcal{E}_0$ represents the edge attributes encoder. Both functions  $\epsilon^{\mathrm{v}}$ and $\epsilon^{\mathrm{e}}$ are implemented as Multi-Layer Perceptrons (MLPs), with 6 hidden layers each. The node encoder input contains 27 features, as described in the previous paragraph, while the edge encoder consists of 4 input features. The dimensions of the latent spaces $\bold{v}_{i,0}$ and $\bold{e}_{j,0}$ are determined by means of a Bayesian hyperparameter optimization scheme, which resulted in a size of 128 latent variables for each one of the encoders.

\subsubsection{Processor}
\label{subsub:processor}

The \textit{processor} is composed of the core block GN$_{\mathrm{core}}$, which consists of $M$ concatenated sub-blocks that are responsible for the information spreading to the nodes. Each one of these sub-blocks is essentially a message-passing step that diffuses the information across the graph and is implemented as an MLP. As such, the quality of the dynamics prediction is affected by the number of message-passing steps, with more complex dynamics usually calling for more steps \cite{learn2simul}. The number $M$ of blocks used in GN\textsubscript{core} along with their corresponding dimensions are also treated as hyperparameters, whose values are obtained from the solution of the hyperparameter optimization problem. The message passing steps are essentially represented by the node and edge update functions $\phi^{\mathrm{e}}$ and $\phi^{\mathrm{v}}$ respectively, as described in \cref{eq:edge-update,eq:node-update}, which are also implemented as MLPs. The input size of the node MLP is $128\times 3$, while the edge MLP consists of a $128\times 2$ input vector. Both MLPs consist of 6 hidden layers, with 114 units and deliver an output of size 128, which is essentially equal to the size of the latent space.

\subsubsection{Decoder}
\label{subsub:decoder}

The \textit{decoder} is the last block of the graph model and receives as input the latent graph $G_{\mathrm{M}}$, which is delivered as output from the \textit{processor}, and decodes the node attributes to the physical space. This is accomplished through the MLP function $\delta^{\mathrm{v}}: \mathcal{V}_{M}\to\mathcal{V}_{\mathrm{d}}$, which translates the node and aggregated edge attributes into the change of position 
$\Delta\mathbf{p}_i$ at each node, so that $\Delta\mathbf{p}_i = \delta^{\mathrm{v}}(\mathbf{v}_{i,M})$.
It should be reminded that the time superscript is herein omitted for the sake of simplicity, however, this change of position is calculated recursively for each time step, thus denoted by $\Delta\mathbf{p}^{t_{k+1}}$, and represents the change of the wall thickness after each trajectory step of the end effector, so that the final position is retrieved as $\mathbf{p}_i^{t_{k+1}} = \mathbf{p}_i^{t_{k}} + \Delta\mathbf{p}_i^{t_{k+1}}$. The input size of the decoder is equal to the size of the latent node representation, that is 128, which is propagated through 6 hidden layers, connected by ReLU activation functions, in order to deliver the change of position in all 3 space dimensions.

\subsection{Training}
\label{subsec:training}

The GNN model consists of five MLP networks, as described in the previous subsection, resulting in a total of 5M trainable parameters, which are learned using the ground-truth data from the experiments described in \cref{sec:problem-description}. The training phase is carried out with the goal of learning the one-step ahead prediction of the node features, as postulated by \cref{eq:gnn-prediction}. Due to the requirement of the model to perform rolling predictions, in which case the prediction at step $t_k$ is fed back to the model for the prediction at step $t_{k+1}$ and so on, white Gaussian noise is added to input of the model, drawn from $\mathcal{N}(0, \sigma_{\mathrm{v}} = 0.003)$. This additive noise results in a reduced accumulation of error across prediction steps as the model learns to handle erroneous predictions. 

The training of the model is based on the use of the augmented dataset, as described in \cref{tab:data-augmentation}, which consists of five different experiments, each of which comprises the printing of 16 layers. 
The training is performed using the Adam optimizer, using a mini-batch gradient descent algorithm with a a maximum of 10.000 gradient steps. As shown in the results section, this is a sufficient number of gradient updating steps for the training. The mini-batch training approach was based on the random sampling of the wall state in terms of node and edge features within a printing trajectory, which was subsequently used for the initialization of the model $\mathcal{M}_{\theta}$ and the prediction of downstream steps. The optimal configuration of the model is obtained from the hyperparameter optimization problem, which is solved using a Bayesian optimization scheme, and resulted in the architecture described in \cref{sub:implementation} for each one of the model components. Lastly, the connectivity radius for wall-to-wall connections is equal to 30mm and the cone radius for wall-trajectory is equal to 0.4$\cdot d$, where $d$ is the wall to end effect distance.

\subsubsection{Loss Function}
\label{subsec:loss-function}

For the training of the GNN model, the loss function is selected in such a way that an overall similarity of the predicted wall thickness with the ground truth is achieved, but also a closer similarity in the region around the spraying gun. To this end, the loss function consists of two terms, as postulated by the following expression

\begin{equation}
    \mathcal{L}(\Delta\hat{\mathbf{p}}^{t_{k+1}};\bm{\theta}) = \lambda_{\Delta}
    \mathcal{L}_{\mathrm{\Delta}}(\Delta\hat{\mathbf{p}}^{t_{k+1}};\bm{\theta}) + \lambda_{\mathrm{HD}}
    \mathcal{L}_{\mathrm{HD}}(\Delta\hat{\mathbf{p}}^{t_{k+1}};\bm{\theta})
\end{equation}

where $\mathcal{L}_{\mathrm{\Delta}}(\Delta\hat{\mathbf{p}}^{t_{k+1}};\bm{\theta})$ denotes the loss term associated with the average predicted position difference between two time steps and $\mathcal{L}_{\mathrm{HD}}(\Delta\hat{\mathbf{p}}^{t_{k+1}};\bm{\theta})$ encodes the maximum distance of the predicted shape with respect to the ground truth. Lastly, the terms $\lambda_{\Delta}$ and $\lambda_{\mathrm{HD}}$ denote the corresponding weights associated with each loss term.

The first loss term aims to minimize the average prediction error in terms of the particles position between successive time steps. This is imposed by the Mean Square Error (MSE), which is further weighted so that a larger weight is assigned to error close to the end effector particle, thus resulting in 

\begin{equation}
    \mathcal{L}_{\mathrm{\Delta}}(\Delta\hat{\mathbf{p}}^{t_{k+1}}; \bm{\theta}) = ||\Delta \hat{\mathbf{p}}^{t_{k+1}} - \Delta \mathbf{p}^{t_{k+1}}||_{W}^{2}
\end{equation}

where $\Delta\hat{\mathbf{p}}^{t_{k+1}}$ is the prediction obtained from the GNN model $\mathcal{M}_{\theta}$, according to \cref{eq:gnn-prediction}, while $\left\lVert \square \right\rVert_W$ designates a weighted norm, which is defined as $\left\lVert \mathbf{x} \right\rVert_W = \sqrt{\mathbf{x}^{\mathrm{T}} \mathbf{W} \mathbf{x} }$, with $\mathbf{x}$ being a vector and $\mathbf{W}$ denoting a symmetric weight matrix. The weight is herein obtained from a Gaussian function, which is centered at the end effector node and its standard deviation is set equal to the value of radius $R^{t_k}$ used for the connectivity of the wall particles with the end effector.

The second term of the loss function is based on the \textit{Hausdorf\-f distance} \cite{hausdorff_dist}, which essentially measures the distance between two surfaces; the ground truth wall formation described by the point cloud $\mathbf{p}^{t_{k+1}}$ and the one predicted by the GNN model, which is accordingly defined by the particle positions $\hat{\mathbf{p}}^{t_{k+1}}$. This loss term is described by the following expression

\begin{equation}
    \mathcal{L}_{\mathrm{HD}}(\Delta\hat{\mathbf{p}}^{t_{k + 1}}; \bm{\theta}) = 
    d(\mathbf{p}^{t_{k+1}}, \mathbf{p}^{t_{k+}}\Delta\hat{\mathbf{p}}^{t_{k+1}}) + 
    d(\mathbf{p}^{t_{k}} + \Delta\hat{\mathbf{p}}^{t_{k+1}}, \mathbf{p}^{t_{k+1}})
    \label{eq:hausdorff-loss}
\end{equation}

where $d(\mathbf{x}, \mathbf{y})$ is the \textit{Hausdorff distance} between the surfaces described by the point clouds $\mathbf{x}$ and $\mathbf{y}$ respectively. This metric measures the maximum distance between two surfaces and is defined as follows

\begin{equation}
    d(\mathbf{x}, \mathbf{y}) = \frac{1}{2} \max_{x\in\mathbf{x}}|x - \text{NN}(x, \mathbf{y})|
    \label{eq:hausdorff-distance}
\end{equation}

where NN denotes the output of the Nearest Neighbor (NN) algorithm. By definition, the \textit{Hausdorff distance} defined in \cref{eq:hausdorff-distance} is not symmetrical, meaning that $d(\mathbf{x}, \mathbf{y}) \neq d(\mathbf{y}, \mathbf{x})$, which results in the symmetric expression adopted in \cref{eq:hausdorff-loss} that is based on both forward and backward distances.

\section{Results}
\label{sec:results}

The results presented in this section aim at showing the performance of the model in comparison to the reference results presented in \cite{plastering}, using an existing benchmark model. This comparison is based on the accuracy in predicting the wall thickness by looking in the one-layer ahead prediction, in which the ground truth data are used for the initialization of the model. Thereafter, the prediction generated at each time step $t_k$ is fed into the model as input for the next step prediction until the entire layer printing is completed. This is considered to be the smallest prediction horizon due to the fact that the actual wall thickness can be measured and used as starting point only at the end of each trajectory of the robotic arm, which corresponds to the printing of a single layer.

\begin{figure}[h]
    \centering
    \begin{subfigure}{0.4\textwidth}
        \includegraphics[width=\textwidth]{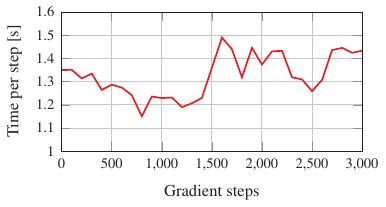}
        \caption{}
        \label{fig:training-time}
    \end{subfigure}
    \begin{subfigure}{0.4\textwidth}
        \includegraphics[width=\textwidth]{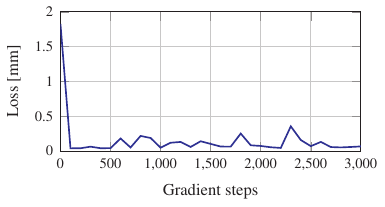}
        \caption{}
        \label{fig:training-loss}
    \end{subfigure}
    \caption{Evolution of the (a) training time and (b) the loss function per gradient step}
    \label{fig:training-time-and-loss}
\end{figure}

\begin{figure}[h]
    \centering
    \includegraphics[width=0.535\textwidth]{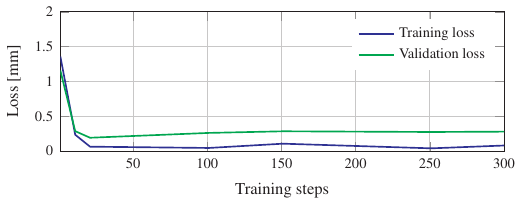}
    \caption{Evolution of training and validation loss during the training phase}
    \label{fig:loss_img_pdf}
\end{figure}

The predictions using the GNN model are generated after training the model with a mini-batch size of 1, resulting to the learning curve presented in \cref{fig:training-time-and-loss}, which corresponds to an average computational time of 1.3s for each update step of the model parameters. The maximum number of gradient updating steps was initially set to 10k, however, as can be observed in the learning curve shown in \cref{fig:training-loss}, the learning requires a significantly smaller amount of steps. As such, the model parameters obtained after 300 gradient steps were selected as the optimal solution, with the learning curve corresponding to these steps shown in \cref{fig:loss_img_pdf}.

\begin{table}[b]
    \centering
    \footnotesize{
    \begin{tabular}{lccccccccccc}
        \toprule
        \multirow{2}{*}{Experiment} & 
        \multirow{2}{*}{\shortstack[c]{Printing \\ area [m\textsuperscript{2}]}} & 
        \multirow{2}{*}{\shortstack[c]{N. of \\ particles}} & 
        \multirow{2}{*}{\shortstack[c]{N. of \\ layers}} & 
        \multirow{2}{*}{\shortstack[c]{Trajectory \\ steps}} & 
        \multirow{2}{*}{\shortstack[c]{Rollout \\ steps}} & & \multicolumn{4}{c}{Velocity [m/s]}\\ \cmidrule{8-11}
        &  &  &  &  &  & & mean & std & max & min \\ \midrule
        S-shaped & 1.34 & 2395 & 12 & 278 & 3336 & & 0.619 & 0.235 & 1.000 & 0.100 \\
        Thunder-shaped &  2.43 & 4290 & 12 & 214 & 2568 & & 0.638 & 0.272 & 1.000 & 0.100 \\
        Wave-shaped &  2.45 &  4364 &  12 & 529 & 6348 & & 0.668 & 0.281 & 1.000 & 0.100 \\
        U-shaped & 1.68 & 3004 & 12 & 445 & 5340 & & 0.595 & 0.158 & 0.983 & 0.326 \\
        \bottomrule
    \end{tabular}
    }
    \caption{Description of the laboratory experiments}
    \label{table:experiments-domain}
\end{table}

The assessment of the model performance is based on four different experiments, with each one corresponding to a different formation. Each one of these experiments, whose names are listed in \cref{table:experiments-domain}, serves as a different dataset for the evaluation of the predictive GNN model ability. All experiments consist of 12 layers, with each one containing different numbers of trajectory steps followed by the spraying gun. Among them, the S-shaped and Thunder-shaped experiments contain the smallest amount of trajectory steps. On the other hand, the Wave-shaped test contains the largest amount of rollout predictions, namely 6348 steps, in combination with the largest domain in terms of area and the highest resolution in terms of the number of particles.

Each experiment is carried out using different operational conditions in terms of the distance to the wall and the printing velocity. The statistics of the latter for each experiment are presented in \cref{table:experiments-domain}. Despite the availability of thickness data from all four experiments, the predictions delivered from the benchmark model are available only for the S-shaped experiment. To this end, a more extensive discussion is provided for the S-shaped experiment, with the prediction results compared not only to the ground truth data but also to the current benchmark performance, which is thoroughly documented in \cite{plastering}.

\begin{figure}[t]
    \centering
    \includegraphics[width=0.775\textwidth]{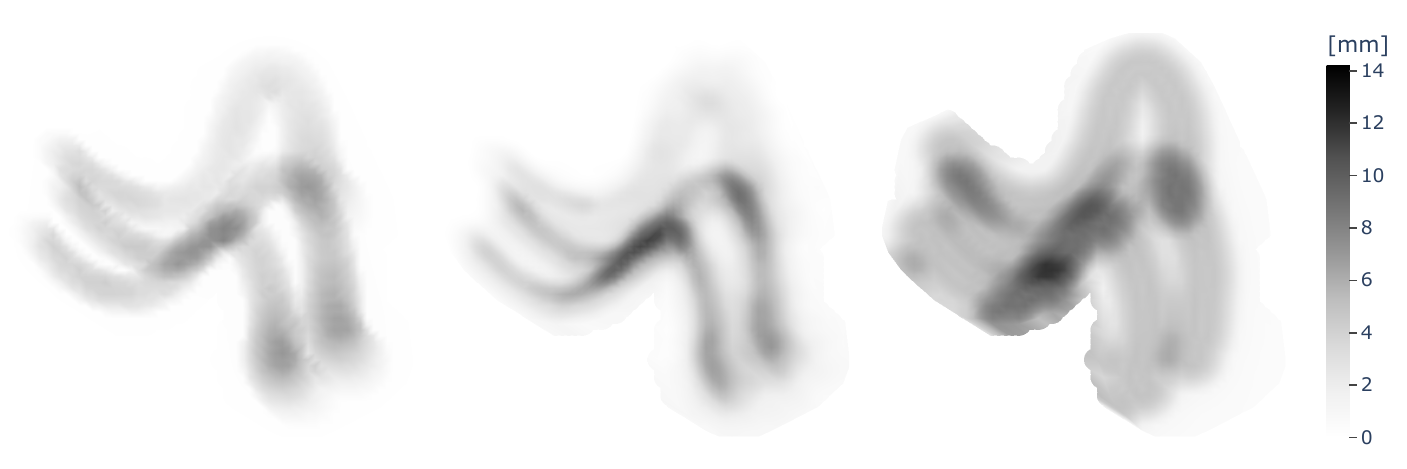}
    \includegraphics[width=0.775\textwidth]{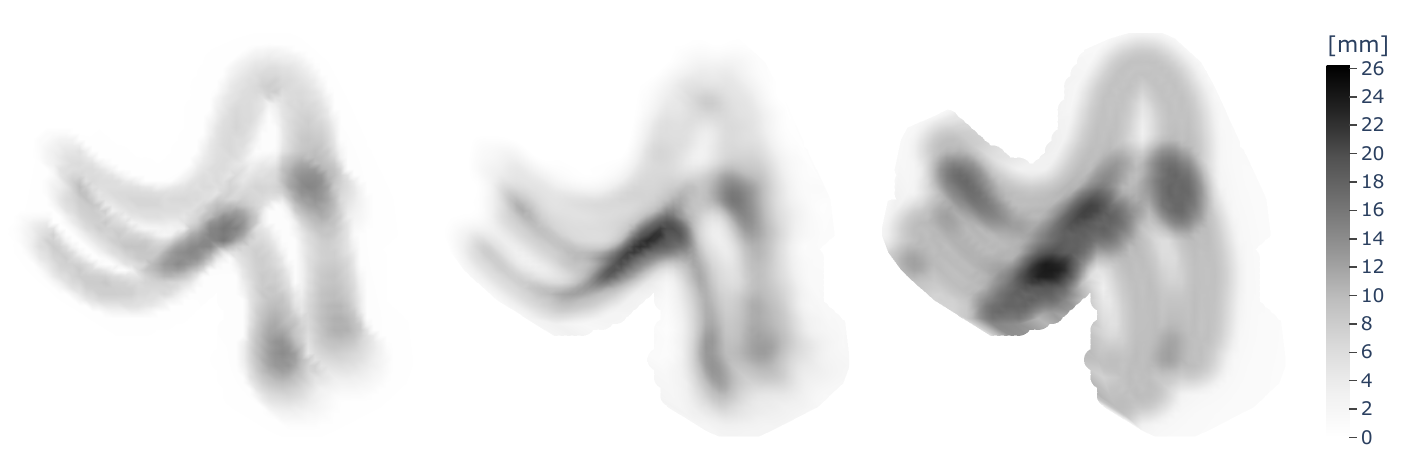}
    \caption{Three- (top) and six-layer (bottom) ahead predictions for the S-shaped experiment; left figure shows the GNN-based thickness prediction, middle figure represents the ground truth data and right figure depicts the thickness predicted by the benchmark model}
    \label{fig:three-and-six-layers-ahead}
\end{figure}

Due to the lack of available measurements between the printing of consecutive layers, the thickness at the end of each layer cannot be used as feedback for correcting the accumulated prediction errors by means of a sequential approach \cite{Tatsis2022}. Consequently, the model's predictive performance is initially tested over a long predictive horizon, that is, for all 12 layers, which is the maximum possible. The results presented in \cref{fig:three-and-six-layers-ahead} show the predictive performance of the model in terms of the three- and six-layer ahead predictions of the S-shaped experiment, which are displayed in the first and second rows respectively. According to the experiments summary presented in \cref{table:experiments-domain}, each printing trajectory of the S-shaped experiment consists of 278 steps, thus resulting in 834 and 1668 rollout predictions respectively. 

\begin{figure}[h]
    \centering
    \includegraphics[width=0.775\textwidth]{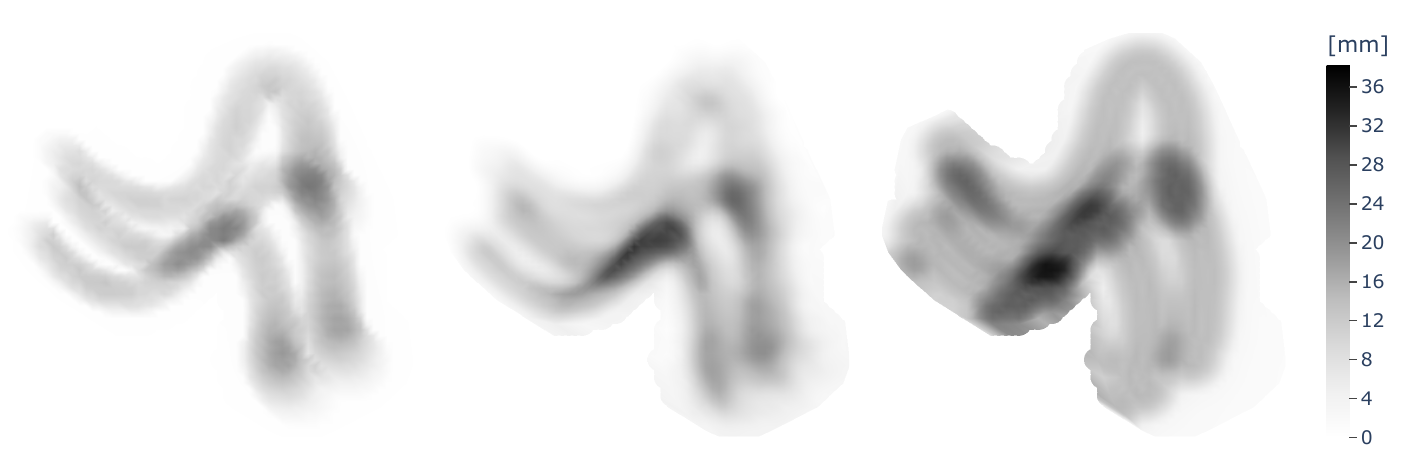}
    \includegraphics[width=0.775\textwidth]{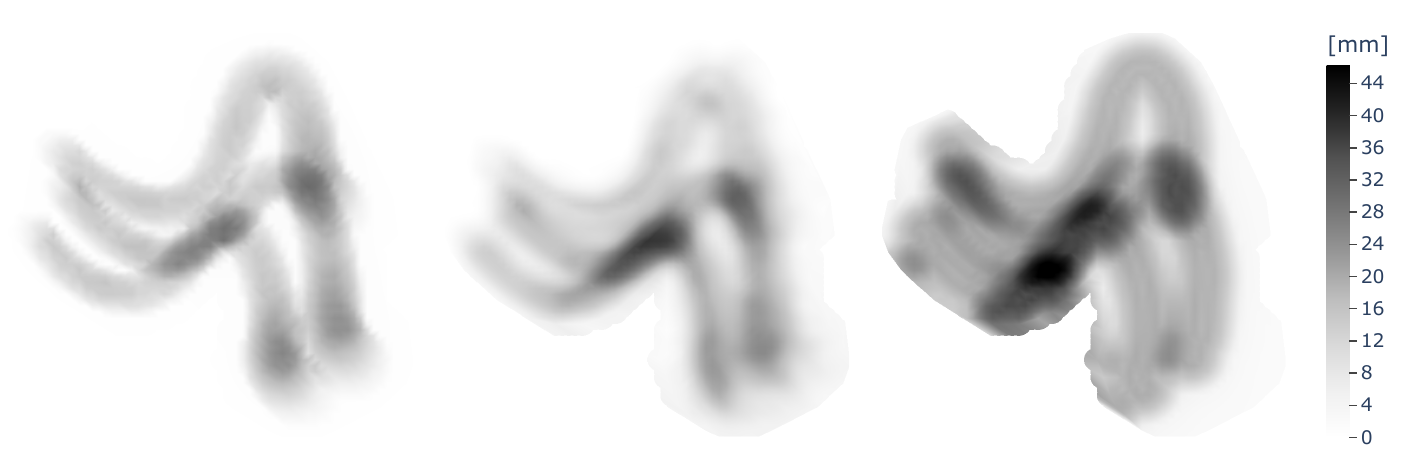}
    \caption{Nine- (top) and twelve-layer (bottom) ahead predictions for the S-shaped experiment; left figure shows the GNN-based thickness prediction, middle figure represents the ground truth data and right figure depicts the thickness predicted by the benchmark model}
    \label{fig:nine-and-twelve-layers-ahead}
\end{figure}

Accordingly, an additional qualitative comparison of the model performance with the ground truth data and the benchmark model is shown in \cref{fig:nine-and-twelve-layers-ahead}, in terms of the nine- and twelve-layers ahead predictions for the S-shaped experiment. It should be noted that the GNN model output is obtained in the form of rollout predictions, resulting in 2502 and 3336 recursive model evaluations, while the output of the benchmark model is delivered on a layer-per-layer basis, thus resulting in 9 and 12 recursive prediction steps respectively.

\begin{figure}[h]
    \centering
    \begin{subfigure}[t]{0.335\textwidth}
        \includegraphics[width = \textwidth]{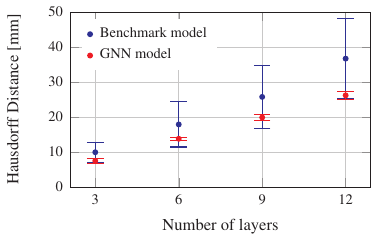}
    \end{subfigure}
    \hspace{10mm}
    \begin{subfigure}[t]{0.335\textwidth}
        \includegraphics[width = \textwidth]{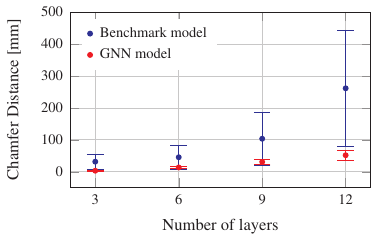}
    \end{subfigure}\vspace{2mm}
    \begin{subfigure}[t]{0.335\textwidth}
        \includegraphics[width = \textwidth]{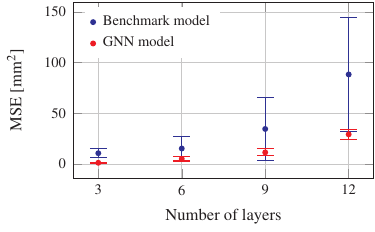}
    \end{subfigure}
    \hspace{10mm}
    \begin{subfigure}[t]{0.335\textwidth}
        \includegraphics[width = \textwidth]{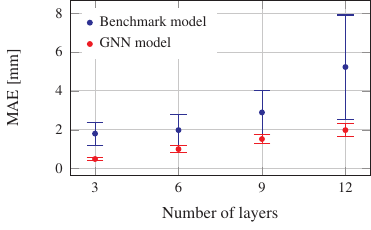}
    \end{subfigure}
    \caption{Comparison of the error metrics between the proposed GNN model and the benchmark model for different prediction horizons of the S-shaped test.}
    \label{fig:error-metrics}
\end{figure}

A quantitative assessment of the results presented in \cref{fig:three-and-six-layers-ahead,fig:nine-and-twelve-layers-ahead} is shown in \cref{fig:error-metrics} using four different error metrics. Namely, the prediction error of the GNN and benchmark models is calculated in terms of the Hausdorff Distance (HD), as introduced in \cref{eq:hausdorff-distance}, the Chamfer Distance (CD), which is also a measure of similarity between two point clouds, as well as the Mean Square Error (MSE) and the Maximum Absolute Error (MAE). It can be seen that the proposed GNN-based modeling approach outperforms the benchmark model and this difference in performance is visible across all four error metrics. Moreover, a consistently increasing offset is observed when looking at change of error metrics from lower to higher prediction horizons. This is due to the fact that both models are employed as recursive predictors, thus resulting in the propagation of errors across sequential prediction steps. It is observed though that each model is characterised by a different scaling of the error, with the GNN model error scaling linearly with respect to the number of predictions steps, while the benchmark model error is scaled exponentially.

\begin{figure}[h]
    \centering
    \includegraphics[width=0.87\textwidth]{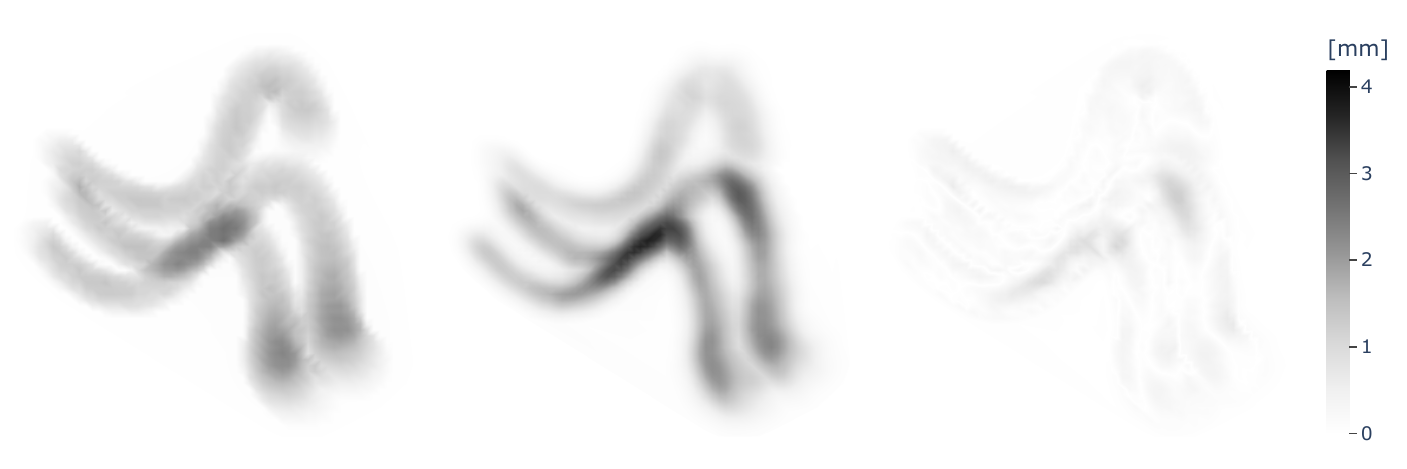}\vspace{-2mm}
    \includegraphics[width=0.87\textwidth]{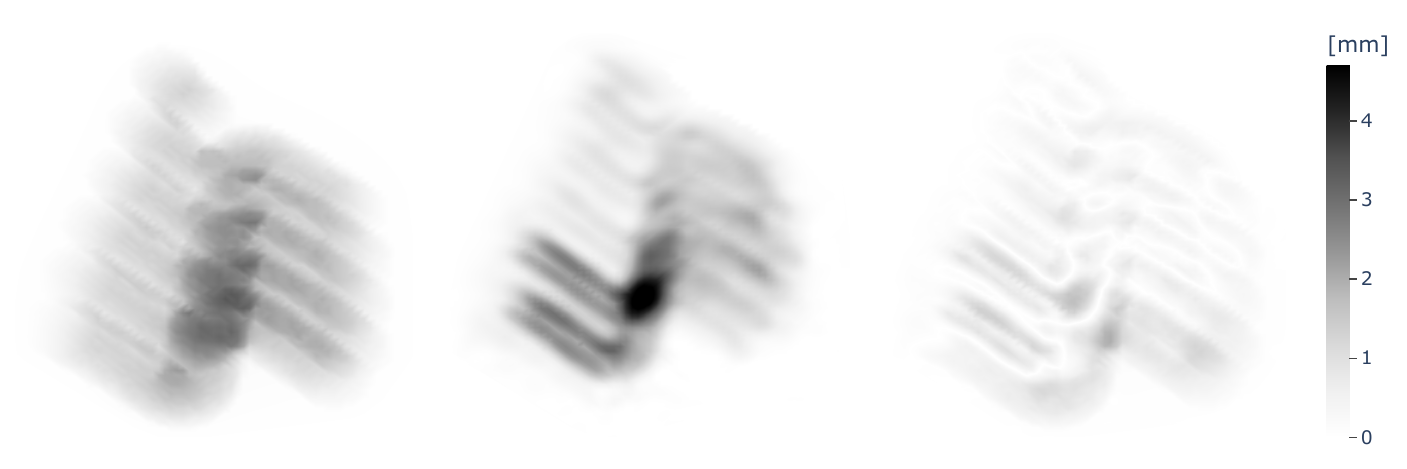}\vspace{-1mm}
    \includegraphics[width=0.87\textwidth]{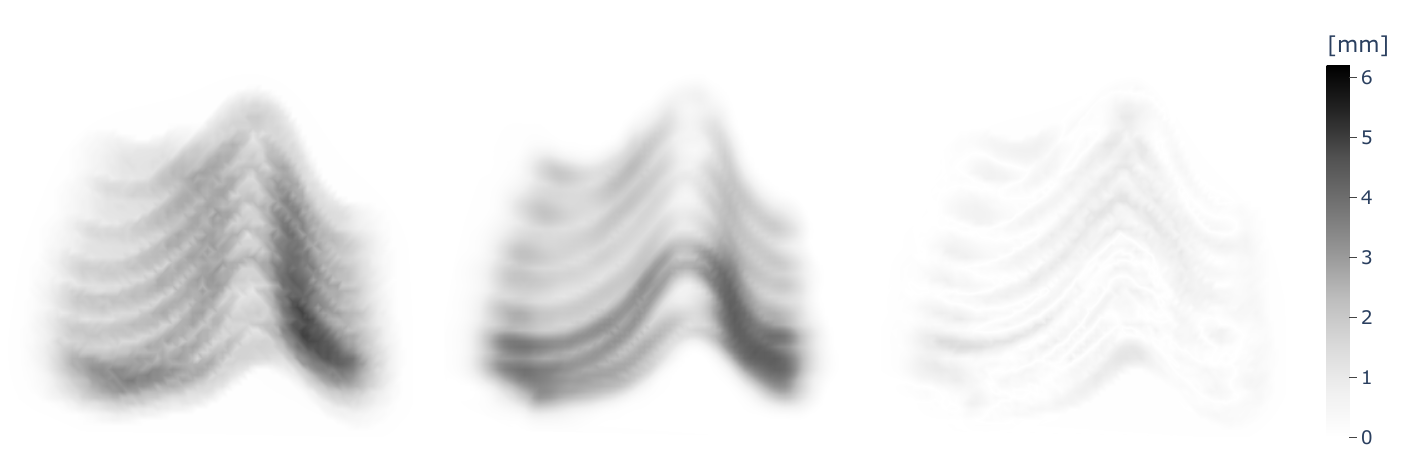}\vspace{1mm}
    \includegraphics[width=0.87\textwidth]{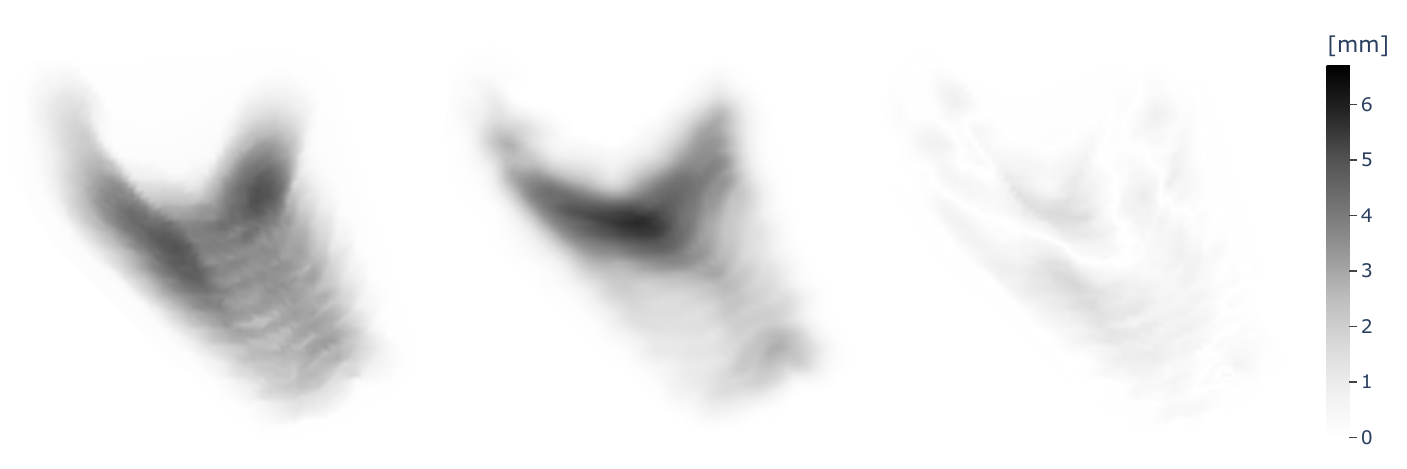}
    \caption{One-layer ahead prediction for the S-, Thunder-, Wave- and U-shaped experiments; left-column figures show the predicted formations, middle-column figures represent the ground truth data and right-column figures depict the absolute error}
    \label{fig:all-four-experiments}
\end{figure}

In order to further assess the ability of the model to generalize its predictive performance, the datasets of all four experiments are used for predictions. It should be noted that the model has been trained with the data presented in \cref{tab:data-statistics}, which correspond to completely different experiments. As such, the data used below for the performance assessment are not seen by the model during training. A qualitative assessment of the one-layer ahead predicted thickness across all four experiments is presented in \cref{fig:all-four-experiments}. Each row of \cref{fig:all-four-experiments} corresponds to a different experiment, while the leftmost column of plots represents the GNN-model predictions, the central plots indicate the ground truth data and the rightmost column of plots contains the absolute error. As discussed in the previous paragraph, the one-layer ahead prediction is an indicative measure of the performance of the model since the error is thereafter linearly accumulated. 

\begin{figure}[h]
    \centering
    \begin{subfigure}[t]{0.365\textwidth}
        \includegraphics[width=\textwidth]{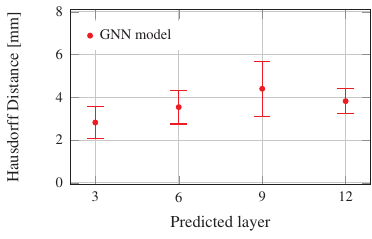}
    \end{subfigure}
    \hspace{10mm}
    \begin{subfigure}[t]{0.365\textwidth}
        \includegraphics[width = \textwidth]{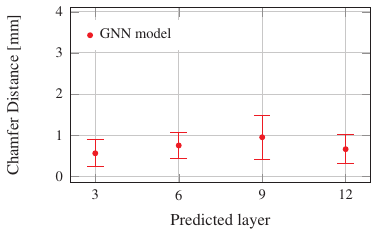}
    \end{subfigure}\vspace{2mm}
    \begin{subfigure}[t]{0.365\textwidth}
        \includegraphics[width = \textwidth]{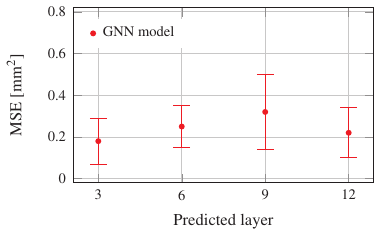}
    \end{subfigure}
    \hspace{10mm}
    \begin{subfigure}[t]{0.365\textwidth}
        \includegraphics[width = \textwidth]{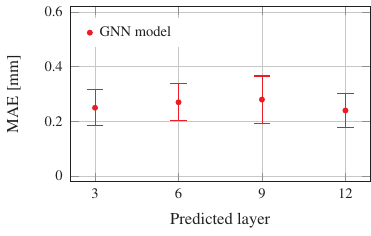}
    \end{subfigure}
    \caption{Error metrics of the GNN-based one-layer ahead predictions for all four tests}
    \label{fig:error-plots-all-experiments}
\end{figure}

A quantitative and extended version of the results shown in \cref{fig:all-four-experiments} is presented in \cref{fig:error-plots-all-experiments} in terms of the Hausdorff distance, the chamfer distance, the mean square error and the maximum absolute error. The error is calculated for the one-layer ahead prediction of layers 3, 6, 9 and 12. This implies that the model state at the end of layer 2 is used for the prediction of layer 3, the state at the end of layer 5 is used for the prediction of layer 6 and so forth. The values of the box plots shown in \cref{fig:error-plots-all-experiments} are calculated as the statistics of all four experiments and it can be seen that the values of all four error metrics are proportionally similar to the those presented in \cref{fig:error-metrics} for the S-shaped test.

\section{Summary}
\label{sec:conclusion}


In this contribution, we introduced a Graph Neural Network-based model for predicting material accumulation on walls printed by a spray-based robotic arm. The proposed modeling approach relies on a particle representation of the problem domain, which enables the adoption of a graph-based solution and the encoding of the printing process and its effect on the wall by means of GNNs. To do so, the available dataset has been augmented in order to enable the learning of a recursive predictor on the basis of time history data that represent the domain thickness at each step of the printing trajectory.

The proposed modeling approach has achieved high accuracy in predicting the target wall formations, not only globally but also in terms of more local features. Moreover, it significantly outperforms the existing benchmark, as demonstrated by all error metrics used in our evaluation. By predicting wall thickness at each trajectory step rather than layer-wise, as carried out by the benchmark model, our model provides a more precise representation of material deposition. Moreover, the data augmentation step, which transforms the available layer-based datasets into trajectory-step-based data enables the integration of our model into a trajectory generation and parameter optimization framework for improved robotic printing applications.

Beyond achieving lower overall error, our model also demonstrates a significant improvement in the scaling of the error with respect to the number of prediction steps. This implies that as predictions progress along the trajectory, the deviation from the ground truth is linearly scaled for our model while it is exponentially scaled when using the benchmark model. This suggests greater robustness and reliability for long-range trajectory predictions. The advancements presented in this work contribute to more accurate and scalable modeling of material deposition, offering valuable applications in automated plastering, trajectory planning and online optimization of the robotic arm and spray gun parameters.

Despite the effectiveness of the proposed GNN approach, there are a few limitations in this version of the model, which are seen as future implementations for delivering more robust and reliable predictions. Namely, the effect of gravity has not been taken into account in the current model, as well as the pressure of the spraying gun, the material density and the spraying angle. The effects of all these parameters has been kept constant during the experiments, without any variability contained in the available dataset. Within this context, one of the future directions consists in the experimental exploration of the entire operational space, which would deliver a more informative and representative dataset. Lastly, the authors aim to further explore the collection of continuous data during the printing process, as well as the generation of a benchmark dataset for the validation of the trained model.

\section{Acknowledgements}
\label{sec:acknowledgements}

Robotic Plaster Spraying was a PhD research project conducted at the Gramazio Kohler Research group at ETH Zurich. It was partially supported by the Swiss National Science Foundation (SNF), within the National Centre of Competence in Research Digital Fabrication (NCCR DFAB
Agreement No. 51NF40-141853), and by the HILTI group with contributions from Giovanni Russo AG. The data used was primarly produced within the Robotic Plaster Spraying project with additions from the follow-up project Robotic On-Site Plastering (ROSP), led by Eliott Sounigo, funded by Innosuisse (Swiss Innovation Agency). Further development efforts on the topic continue within the newly founded start-up company LAYERED.

\bibliographystyle{unsrt}
\bibliography{references}  

\end{document}